\pdfoutput=1
\RequirePackage[T1]{fontenc}
\documentclass[12pt]{article}

\usepackage[height=8.85in,width=6.45in]{geometry}

\usepackage[utf8]{inputenc}
\usepackage{amsmath}
\usepackage{amssymb}
\usepackage{mathtools}
\numberwithin{equation}{section}
\usepackage{slashed}
\usepackage{braket}
\usepackage[svgnames]{xcolor}
\usepackage[colorlinks,citecolor=DarkGreen,linkcolor=FireBrick,linktocpage]{hyperref}
\usepackage{cite}
\usepackage{color}
\usepackage{tikz}
\usepackage{tikz-cd}
\usepackage{times}
\usepackage{courier}
\usepackage{bm}
\usepackage{subfig}
\usepackage{authblk}
\usepackage[pgf]{adjustbox}

\allowdisplaybreaks

\def\bZ{\mathbb{Z}}
\def\bR{\mathbb{R}}
\def\bC{\mathbb{C}}
\def\cA{\mathcal{A}}
\def\cC{\mathcal{C}}

\def\cM{\mathcal{D}}
\def\cM{\mathcal{M}}
\def\sT{\mathsf{T}}

\def\RP{\mathbb{RP}}
\def\CP{\mathbb{CP}}
\def\CC{\mathbb{CC}}
\def\MO{\mathbb{MO}}
\def\KB{\mathbb{KB}}
\def\sP{\mathsf{P}}
\def\rP{\mathrm{P}}
\def\rT{\mathrm{T}}
\def\rC{\mathrm{C}}
\def\rS{\mathrm{S}}

\def\non-orientable{\text{non-orientable}}

\def\Ker{\mathop{\text{Ker}}}
\def\Im{\mathop{\text{Im}}}

\def\tr{\mathop{\text{Tr}}}
\def\dim{\mathop{\text{dim}}}

\def\underline#1{\bm{#1}}

\begin{document}

\begin{titlepage}

\title{On dimensions of (2+1)D abelian bosonic topological systems on non-orientable manifolds 
}
\author{Ippo Orii}
\affil{Kavli Institute for the Physics and Mathematics of the Universe, \\
University of Tokyo, Kashiwa, Chiba 277-8583, Japan}

\date{\empty} 

\maketitle

\begin{abstract}
We give a framework to describe abelian bosonic topological systems with parity symmetry on a torus 
in terms of the projective representation of $GL(2,\bZ)$. 
However, this information alone does not guarantee that we can assign Hilbert spaces to non-orientable surfaces
in a way compatible with the gluing axiom of topological quantum field theory.
Here, we show that we may assign Hilbert spaces with integer dimensions to non-orientable surfaces
in the case of abelian bosonic topological systems with time-reversal symmetry, which can be seen as a necessary condition for the existence of topological quantum field theories.

\end{abstract}

\end{titlepage}

\tableofcontents

\section{Introduction and summary}
Topological quantum field theories (TQFTs) in 2+1 dimensions have been studied for three decades as an interactive area of particle physics and topology. They are also used to give low-energy descriptions of two-dimensional gapped systems in condensed matter physics. For example, the fractional quantum Hall effect can be described in terms of TQFTs.

When constructing 2+1 dimensional TQFTs, there are two general approaches. One way is to start from Atiyah's axioms to analyze geometrically\cite{Atiyah:1989vu}, which is our standpoint and mainly discussed in this paper. Another way is to describe the theory in terms of the modular tensor category (MTC) to analyze algebraically\cite{Moore:1988qv}.
Those two ways are known to have correspondence with each other.

In many cases, we assume that the space-time manifolds are oriented in TQFTs. Recently, however, it has become necessary to work on TQFTs without orientability, such as in the case where we work on symmetry protected topological (SPT) phases with time-reversal symmetry\cite{PhysRevB.75.121306,RevModPhys.89.040502}. Much remains to be discovered regarding TQFTs on non-orientable manifolds. The correspondence between the two constructions is not fully understood in such cases.

In 2+1 dimensional theories, we often work on the Hilbert space on a torus. Oriented TQFTs assign a Hilbert space to a given oriented closed surface, and every oriented connected closed surface is homeomorphic to a two-sphere or the connected sum of $g$ tori. (See e.g. \cite[Sec.~6.4]{book}.) Therefore, the analysis on a torus is fundamental for understanding the general cases.

Regarding the torus, the Hilbert space on $T^2$ is known to be the representation space of $SL(2,\bZ)$\cite{Kitaev_2006}. However, when it comes to non-orientable cases, we did not have an explicit way to describe the theories in terms of a representation of a group.
Therefore, we introduce a $P$ matrix for abelian bosonic anyons and show that it forms the projective representation of $GL(2,\bZ)$ with $S,T,C$ matrices already known. 

Furthermore, there is another problem with non-orientable theories as to the dimension. The dimension can be computed using the formula $\mathop{\text{dim}}V(\Sigma)=Z(\Sigma\times S^1)$, where $\Sigma$ is a two-dimensional oriented closed surface, $V(\Sigma)$ is a Hilbert space on it, and $S^1$ is a one-dimensional sphere.

This formula is defined on oriented TQFTs and known to give an integer, which might seem trivial. But it is not clear in non-orientable cases. In fact, from the point of view of MTC, it is considered one of the conditions which must be satisfied  for the triviality of the \(H^3\) obstruction\cite{Barkeshli:2017rzd}. In other words, this can be regarded as a necessary condition for the existence of a TQFT on non-orientable manifolds. Throughout this discussion, we have implicitly adopted the perspective of the Reshetikhin--Turaev construction. At present, however, there is no concrete construction of a Reshetikhin--Turaev TQFT applicable to non-orientable manifolds in full generality. Consequently, the failure of $M_a$ to be an integer should not be viewed as a definitive inconsistency, but rather as an indication that the given MTC data cannot be straightforwardly extended to a consistent TQFT on non-orientable manifolds, at least within this framework.

     As a step for general cases, we analyze the abelian bosonic type of TQFTs. In that case, we show that we can compute the dimension without problem even in non-orientable cases, which is the main result of this paper. Note that our discussion is mainly on $\RP^2$. Since we can have any non-orientable surface by a connected sum using $\RP^2$, it depends only on $\RP^2$ whether the dimension is an integer in general. In the discussion, we introduce a set $\cM$ as a subset of the whole anyon group, which gives nonzero value to the dimension of $\RP^2$ when inserted. We derive several facts on $\cM$, and using it to show one constraint which must be satisfied in terms of MTC, discussed previously for example in \cite[Sec.~II.E]{Barkeshli:2017rzd} and \cite[Sec.~2]{Lee:2018eqa}. Furthermore, we derive an explicit form of $\cM$ and get some constraints associated with it.

In addition, we review the discussion in  \cite[Eq.~(6)]{Wang_2017} and \cite[Sec.~VII]{Barkeshli:2016mew} which is related to our result and the time-reversal anomaly.
 In general, 2+1 dimensional abelian bosonic systems have a $\bZ_2\times\bZ_2$ classification. And they are characterized by two phases given by $Z(\RP^4)$ and $Z(\CP^2)$, which are the partition functions of the corresponding 3+1 dimensional SPT phases\cite{kapustin2014symmetryprotectedtopologicalphases}. We see that the product of the two values turns out to be the topological spin of the anyon in $\cM$.

\paragraph{Organization of the paper :}
In Sec.~2, we give a quick review of Atiyah TQFT and analyze the crosscap state. The coefficients of the crosscap state, which we denote as $M_a$, can be considered as the dimension of the Hilbert space on $\RP^2$ with a line operator $a$ inserted.
In Sec.~3, we give a set of data of abelian bosonic systems and construct a projective representation of $GL(2,\bZ)$.
In Sec.~4, we derive the explicit form of $M_a$, which leads to our conclusion that the dimension of the Hilbert space can be computed in abelian bosonic systems on any non-orientable surface. In addition, we discuss related facts on $M_a$ such as the consistency condition and time-reversal anomaly. In Sec.~5 we summarize our results and comment on future work.

\section{Topological analysis}
\subsection{Atiyah's Axioms of TQFT}
Here, we quickly review axioms of TQFT by Atiyah \cite{Atiyah:1989vu}.
One can skip the whole section if he or she is already familiar with the concepts.
\paragraph{Axiom I :}A $D$-dimensional space $\Sigma$ is associated with a Hilbert space $V(\Sigma)$ which depends only on the topological structure of $\Sigma$. 

\paragraph{Axiom II :}The disjoint union of two $D$-dimensional spaces $\Sigma_1$ and $\Sigma_2$ will be associated with a Hilbert space which is the tensor product of the Hilbert spaces associated with each space. I.e.,
\begin{equation}
    V(\Sigma_1\cup\Sigma_2)=V(\Sigma_1)\otimes V(\Sigma_2).
\end{equation}

This implies the next property:
\begin{equation}
    V(\varnothing)=\mathbb{C}.
\end{equation}

\paragraph{Axiom III :}If $M$ is a $(D+1)$-dimensional manifold with $D$-dimensional boundary $\Sigma=\partial M$, then we associate a particular element of the vector space $V(\Sigma)$ with this manifold, which we denote as
\begin{equation}
    Z(M) \in V(\partial M).
\end{equation}

This means that if we take the argument manifold $M$ to be an closed one, we have
\begin{equation}
    Z(M) \in V(\partial M) = V(\varnothing) = \mathbb{C}.
\end{equation}

\paragraph{Axiom IV :}Reversing orientation is denoted as $V(\Sigma^\ast)=V(\Sigma)^\ast$ where by $\Sigma^\ast$ we mean the same surface with reversed orientation, whereas by $V^\ast$ we mean the dual space.

Now, from these axioms, we construct tools to describe TQFTs. 
\paragraph{Gluing :}Suppose we have two manifolds $M$ and $M'$ with a common boundary $\Sigma$ but opposite orientation. Then, we can consider an inner product by gluing the two manifolds along the boundary $\Sigma$.
\begin{equation}
\begin{aligned}
    Z(M)&\coloneqq \ket{\psi}\in V(\Sigma),\\
    Z(M')&\coloneqq \bra{\psi '}\in V(\Sigma)^\ast,\\
     Z(M\cup_{\Sigma}M')&\coloneqq\braket{\psi'|\psi}\in V(\varnothing)=\mathbb{C}.
\end{aligned}
\end{equation}
Here, we mean that $M$ and $M'$ are glued together along $\Sigma$ by $M\cup_{\Sigma}M'$.

\paragraph{Cobordism :}Suppose that the disjoint union of $\Sigma_1$ and $\Sigma_2$ are the boundary of a manifold $M$. Then, $M$ is called a cobordism between $\Sigma_1$ and $\Sigma_2$, and this concept gives us the method of transformations in TQFTs. Considering Axiom III and IV, we have
\begin{equation}
    \begin{aligned}
        \partial M&= {\Sigma_1}^\ast \cup \Sigma_2,\\
         Z(M)&\in V(\Sigma_1)^\ast\otimes V(\Sigma_2).
    \end{aligned}
\end{equation}
Therefore, we can write $Z(M)$ as 
\begin{equation}
Z(M)=\sum_{a,b}U_{ab}\ket{\psi_{{\Sigma_2},a}}\otimes\bra{\psi_{{\Sigma_1},b}}.
\end{equation}
Here, we denote the basis of each Hilbert space as $\ket{\psi_{{\Sigma_2},a}}$ and $\bra{\psi_{{\Sigma_1},b}}$.

\paragraph{Projective phase :}Using the concept of cobordism, we can consider a successive transformation $V(\Sigma_1)\rightarrow V(\Sigma_2)$, then $V(\Sigma_2)\rightarrow V(\Sigma_3)$, which results in $V(\Sigma_1)\rightarrow V(\Sigma_3)$. For example, suppose we have two manifolds $M$ and $M'$ as follows:
\begin{equation}
    \begin{aligned}
        \partial M&= {\Sigma_1}^\ast \cup \Sigma_2,\\
         Z(M)&\in V(\Sigma_1)^\ast\otimes V(\Sigma_2),\\
         \partial M'&= {\Sigma_2}^\ast \cup \Sigma_3,\\
         Z(M')&\in V(\Sigma_2)^\ast\otimes V(\Sigma_3).
    \end{aligned}
\end{equation}
Let us consider the manifold $N=M\cup_{\Sigma_2}M'$ given by gluing $M$ and $M'$ along one of their boundaries $\Sigma_2$. Then we have
\begin{equation}
    \begin{aligned}
         \partial N&= {\Sigma_1}^\ast \cup \Sigma_3,\\
         Z(N)&\in V(\Sigma_1)^\ast\otimes V(\Sigma_3).
    \end{aligned}
\end{equation}
Therefore, both $Z(M')Z(M)$ and $Z(N)$ define a map $V(\Sigma_1)\rightarrow V(\Sigma_3)$. We hope that the two maps can be identified, but sometimes we have a projective phase as follows:
\begin{equation}
    Z(N)=\phi(M,M')Z(M')Z(M)
\end{equation}
where $\phi(M,M')\in U(1)$ is the projective phase which depends on $M$ and $M'$. Such situations are called anomalous, and we often have interests in whether the phase is trivial or not. In fact, we discuss the projective phase associated with the reflection acting on $V(T^2)$ in Sec.~3.3.
One approach to take care of this projective phase is to consider our TQFT to be on the boundary of a $(D+2)$-dimensional bulk.
This point of view was taken e.g.~in \cite{Barkeshli:2016mew}.
Here we use a more primitive but easier-to-use formulation with projective phases.

\paragraph{Identity cobordism :}Suppose that $M= \Sigma \times I$, where $I$ is the one dimensional interval. Interval is topologically trivial and can be contracted to nothing, therefore $Z(M)$ is an identity map defined on the Hilbert space associated with $V(\Sigma)$ as follows:
\begin{equation}
    \begin{aligned}
         M&=\Sigma \times I,\\
         \partial M&=\Sigma^\ast \cup \Sigma,\\
        Z(M)&=\sum_a\ket{\psi_{\Sigma,a}}\otimes\bra{\psi_{\Sigma,a}}= \text{identity}.
    \end{aligned}
\end{equation}

With this identity map, we can compute the dimension of the Hilbert space associated with the closed surface $\Sigma$ in the following way:
\begin{equation}
    \begin{aligned}
        \mathop{\text{Tr}}(Z(\Sigma\times I))
        &=\text{Tr}\sum_a\ket{\psi_{\Sigma,a}}\otimes\bra{\psi_{\Sigma,a}}\\
           &=\sum_a \braket{\psi_{\Sigma,a}|\psi_{\Sigma,a}}\\
           &=\mathop{\text{dim}}V(\Sigma).\\
    \end{aligned}
\end{equation}
Here, we take the bases normed to one.
Since we can also consider $\text{Tr}(Z(\Sigma\times I))$ as $Z(\Sigma\times S^1)$, we have
\begin{equation}
    Z(\Sigma\times S^1) = \dim V(\Sigma).\label{dimformula}
\end{equation}

This is the formula to compute the dimension of the Hilbert space, so what we aim to show is that
\begin{equation}
   Z(\Sigma\times S^1)\in\bZ \label{dimformula}
\end{equation}
for any non-orientable two-dimensional closed surface $\Sigma$ in abelian bosonic cases.

\paragraph{Inserting line operators :}Let us introduce the way to give a complete basis of the Hilbert space on a surface. We take $T^2$ as an example of the surface, which will appear in our discussion later. 

First, note that $T^2$ is given as a boundary of $D^2_A\times S^1_B$, where $D^2$ is a two-dimensional disk. Here, we introduce indices $A$ and $B$ to distinguish the two $S^1$. Then, we can insert a line operator corresponding to the world-line of an anyon, and use it for labeling of the bases of the Hilbert space. In this case, we consider $D^2_A(a)\times S^1_B$, where the line operator $a$ is inserted at the center of the disk $D^2_A$ times $S^1_B$. The boundary of this manifold is again $T^2=S^1_A\times S^1_B$. Therefore TQFT assigns an element of $V(T^2)$, and let us denote it as $\ket{a}$:
\begin{equation}
    \ket{a}\coloneqq Z(D^2_A(a)\times S^1_B).\label{basisa}
\end{equation}
The same procedures with every anyon in the system gives us a complete basis of $V(T^2)$.

Next, we consider an exchange of the roles of $A$ and $B$. That means we consider a manifold $S^1_A\times D^2_B(a)$, which again gives us an element of $V(T^2)$. This state is given by the modular transformation $S\in SL(2,\bZ)$ acting on the state we have in \eqref{basisa}:
\begin{equation}
    S\ket{a}\coloneqq Z(S^1_A\times D^2_B(a)).
\end{equation}
The same procedures with every anyon in the system gives us another complete basis of $V(T^2)$.

\paragraph{Why $\RP^2$? :}In order to analyze the dimension of the Hilbert space with some non-orientable surfaces, it is enough to analyze the one with $\RP^2$. Why is that? First, if we get one closed surface by a connected sum of two closed surfaces with integer dimensions, the resulting surface also has an integer dimension. For example, suppose we have two closed surfaces $M$ with anyon $a$ and $M'$  with anyon $b$, and denote each dimension as $N_a$ and $N'_b$, then the dimension of the resulting surface of the connected sum is $\sum_{a,b} N_a\delta^{\overline{a} b}N'_b = \sum_a N_aN'_{\overline{a}}$. It immediately follows that the right-hand side of this equation is an integer if $N_a$ and $N'_a$ are integers.\footnote{See \cite[Sec.~8]{Simon:2023hdq} for the general case.}
Second, any non-orientable (compact) surface is homeomorphic to a connected sum of some projective planes. (See e.g. \cite[Sec.~6.4]{book}.) 
Therefore, we aim to show that the value which is supposed to be the dimension of the Hilbert space on $\RP^2$ with abelian anyons system is an integer: 
\begin{equation}
   M_a \coloneqq Z(\RP^2(a)\times S^1) \in\bZ
   \label{Z(RP2)},
\end{equation}
where $\RP^2(a)\times S^1$ denotes $\RP^2\times S^1$ with a line operator $a$ inserted.\footnote{Of course, if a non-orientable extension of the theory exists, this quantity must be a nonnegative integer, since it is interpreted as the dimension of a Hilbert space. However, strictly speaking, the Reshetikhin--Turaev TQFT is constructed only for orientable manifolds, and there is no general proof that such a TQFT admits an extension to non-orientable manifolds for an arbitrary MTC. Therefore, if one encounters a case with \(M_a\notin\mathbb{Z}\), this should be interpreted as an obstruction: the given MTC data cannot define a consistent non-orientable extension, at least in a naive sense.
}

\subsection{Analysis on the crosscap state}

In this section, by reviewing the discussions in \cite{Barkeshli:2016mew,Tachikawa_2017,Barkeshli:2017rzd}, we give an explicit form of the crosscap state, which is closely related to $M_a$.

Consider the manifold $\MO_A\times S^1_B$, where $\MO_A$ is a Möbius strip, connecting the boundary $S^1_A$ and the crosscap given by
\begin{equation}
    \MO_A=\{(x,\theta)\in[-1,1]\times\bR\mid(x,\theta)\sim(-x,\theta +\pi)\}.
    \label{moa}
\end{equation}
Then we define the crosscap state $\ket{\CC}$ as 
\begin{equation}
    \ket{\CC}=Z(\MO_A\times S^1_B).
\end{equation}
Note that $\partial\MO_A=S^1_A$, then the fact that $\partial(\MO_A\times S^1_B)=S^1_A\times S^1_B=T^2$ allows us to expand the crosscap state by the complete basis we gave in \eqref{basisa}:
\begin{equation}
\ket{\CC}=\sum_{a}\tilde{M}_a\ket{a},
\end{equation}
where we use $\tilde{M}_a$ as coefficients. Now, consider an inner product $\braket{a|\CC}=\bra{a}\sum_{b}\tilde{M}_b\ket{b}=\tilde{M}_a$. What we did here is equivalent to gluing two manifolds $D^2_A(a)^\ast\times S^1_B$ and $\MO_A\times S^1_B$ along the same boundary $T^2$. And note that $\MO_A$ is glued with $D^2_A(a)$ along the same boundary $S^1_A$ and is deformed to $\RP^2(a)$ as a result. Then we have
\begin{equation}
    \tilde{M}_a=Z(\RP^2(a)\times S^1_B)=M_a.
\end{equation}
Therefore we have
\begin{equation}
\ket{\CC}=\sum_{a}M_a\ket{a}.\label{maaa}
\end{equation}

Now we recall that we can span the crosscap state by another basis:
\begin{equation}
    \ket{\CC}=\sum_{a}\eta(a)S\ket{a} ,\label{etasa}
\end{equation}
where we use $\eta(a)$ as coefficients. 
With \eqref{maaa}, \eqref{etasa}, 
we have
\begin{equation}
    M_a=\sum_{b\in\cA}S_{ab}\eta(b) , \label{manotker}
\end{equation}
where $S_{ab}=\bra{a}S\ket{b}$.

Now get back to the crosscap state, the coefficients $\eta(a)$ can be computed as
\begin{equation}
    \eta(a)=\bra{a}S^{-1}\ket{\CC}.
\end{equation}
Let us look at it from the geometrical point of view. We basically follow the discussion in \cite{Barkeshli:2017rzd} and \cite{Tachikawa_2017}. Recall the fact that $S\ket{a}$ corresponds to a manifold $S^1_A\times D^2_B(a)$, then we see that $\bra{a}S^{-1}\ket{\CC}$ can be interpreted as the partition function on the manifold obtained by gluing $\MO_A\times S^1_B$ and $S^1_A\times D^2_B(a)^\ast$ along the boundary $T^2=S^1_A\times S^1_B$. We denote this manifold as $X(a)$.

For now, we forget about the line $a$ so that we focus on how the manifold looks like. First, take the oriented double cover of $\MO_A$ in \eqref{moa} as 
\begin{equation}
    \tilde{\MO}_A=[-1,1]\times S^1_A.
\end{equation}
Then the oriented double cover of $X$, which we denote as $\tilde{X}$, is given by gluing $S^1_A\times D^2_B$ and its reflected one to $\tilde{\MO}_A\times S^1_B=S^1_A\times [-1,1]\times S^1_B$. As a result, we have
\begin{equation}
    \tilde{X}=S^1_A\times S^2_B,
\end{equation}
where $D^2_B$ and its reflected one are glued to $[-1,1]\times S^1_B$, resulting in $S^2_B$. 

We move on to the next step to reduce $\tilde{X}$ to $X$. We describe $S^1_A$ and $S^2_B$ as
\begin{equation}
\end{equation}
For the purpose of the reduction, we introduce a diffeomorphism $\sigma$ acting on $S^1_A\times S^2_B$ as
\begin{equation}
    \sigma:(\theta,n_x,n_y,n_z)\mapsto(\theta+\pi,-n_x,n_y,n_z).
\end{equation}
Then we see that we get $X$ by the quotient of $\tilde{X}=S^1_A\times S^2_B$ by the action of $\sigma$:
\begin{equation}
    X=[S^1_A\times S^2_B]/\sigma.
\end{equation}

Recalling that the line $a$ is inserted at $D^2_B$, we have
\begin{equation}
    \tilde{X}(a)=S^1_A\times S^2_B(a,\mathsf{P}a).
\end{equation}
Here $\sP a$ is the type of line operator which is given by reflecting the line operator $a$, $S^2_B(a,\sP a)$ is the $S^2_B$ with the line operators $a$ and $\sP a$ inserted at the north pole $(n_x,n_y,n_z)=(1,0,0)$ and the south pole $(n_x,n_y,n_z)=(-1,0,0)$, respectively.

In summary, we get
\begin{equation}
    X(a)=[S^1_A\times S^2_B(a,\sP a)]/\sigma
\end{equation}

Let $V(S^2(a,a'))$ be the Hilbert space on $S^2_B$ with the line operators $a$ and $a'$ inserted at the north pole and the south pole, respectively. And let $P_{S^2}$ be the operator $P_{S^2}:V(S^2_B(a,a'))\rightarrow V(S^2_B(\sP a',\sP a))$ which implements $(n_x,n_y,n_z)\rightarrow (-n_x,n_y,n_z)$. Taking $a'$ to be $\sP a$, $P_{S^2}$ is an endomorphism $P_{S^2}:V(S^2_B(a,\sP a))\rightarrow V(S^2_B(a,\sP a))$. Then our geometrical analysis results in the following:
\begin{equation}
    \eta(a)=\tr {}_{V(S^2(a,\sP a))}(P_{S^2}).\label{etatr}
\end{equation}
Here,
\begin{equation}
\begin{aligned}
    \eta(a)&\in U(1)&\text{for all } a\text{ such that }\overline{a}=\sP a,\\
    \eta(a)&=0 &\text{for all } a\text{ such that }\overline{a}\ne\sP a
\end{aligned}
\end{equation}
because $\dim V(S^2(a,\sP a))=\delta_{\overline{a},\sP a}$. Therefore we have 
\begin{equation}
    M_a=\sum_{\overline{b}=\sP b}S_{ab}\eta(b).\label{maa}
\end{equation}
In~\cite{Tachikawa_2017}, \(\eta(a)\) is identified with the local Kramers degeneracy~\cite{Wang_2017}. Accordingly, when \(\overline{a}=\sP a\), it is expected to be interpreted as the symmetry fractionalization data in the MTC~\cite{Barkeshli:2016mew}.\footnote{In order to define symmetry fractionalization, the \(H^3\) obstruction must be trivial. This has been shown to be the case for all abelian bosonic systems in~\cite{Orii:2025hgn}. Therefore, in what follows, we can discuss symmetry fractionalization without concern for the \(H^3\) obstruction.
} For this purpose, we impose the following assumption on the map \(\eta\):
\begin{equation}
    \eta(a+b)=\eta(a)\,\eta(b),
\end{equation}
for all \(a,b\) satisfying \(\overline{a}=\sP a\) and \(\overline{b}=\sP b\).
This multiplicativity is a standard property in the MTC framework.
In Sec.~4.3, we will derive several consistency conditions that must be satisfied in abelian bosonic systems within this MTC setup, including
\(\eta(a)\in\{\pm1\}\) and
\(
    B(a,-\sP a)\,\eta\bigl((1-\sP)a\bigr)=1,
\)
in agreement with the results of~\cite{Barkeshli:2016mew,Wang_2017,Lee:2018eqa}.

The discussion above is just a simple version of the one in \cite[Sec.~2]{Tachikawa_2017}. This result is also derived in \cite[Sec.~VII]{Barkeshli:2016mew}.
We can show another interesting result almost in the same way.

Let us consider the following:
\begin{equation}
    \dim V(\KB)=Z(\KB\times S^1_C),
\end{equation}
where $\KB$ is the Klein bottle and the index $C$ will be needed to distinguish it from other $S^1$ in the following.
Note that $\KB$ is given by a connected sum of two $\RP^2$, then we see that\footnote{Please also see the footnote 1.}
\begin{equation}
    Z(\KB\times S^1_C)=\sum_{a}M_aM_{\overline{a}}. \label{dimkbb}
\end{equation}
By a geometrical analysis, we will find a similar formula to \eqref{etatr}.
First we describe $\KB$ as follows:
\begin{equation}
    \KB=\{(\theta,\phi)\in S^1_A\times\bR \mid(\theta,\phi)\sim(-\theta,\phi+\pi)\},
\end{equation}
where $S^1_A=\{\theta\in\bR\mid\theta\sim\theta+2\pi\}$.
Then, take the oriented double cover of $\KB$ as
\begin{equation}
    \tilde{\KB}=S^1_A\times S^1_B.
\end{equation}
Therefore the oriented double cover of the ordinary manifold is
\begin{equation}
    \tilde{\KB}\times S^1_C=S^1_A\times S^1_B\times S^1_C=S^1_B\times T^2,
\end{equation}
where $T^2=S^1_A\times S^1_C=\{(\theta,\varphi)\in\bR\times\bR\mid\theta\sim\theta+2\pi,\varphi\sim\varphi+2\pi\}$.
We move on to the next step to reduce $\tilde{\KB}$ to $\KB$. For the purpose of the reduction, we introduce a diffeomorphism $\sigma'$ acting on $S^1_B\times T^2$ as
\begin{equation}
    \sigma':(\phi,\theta,\varphi)\mapsto(\phi+\pi,-\theta,\varphi).
\end{equation}
Then we see that
\begin{equation}
    \KB\times S^1_C=[S^1_B\times T^2]/\sigma'.
\end{equation}
Let $P_{T^2}$ be the operator $P_{T^2}:V(T^2)\rightarrow V(T^2)$ which implements $(\theta,\varphi)\mapsto(-\theta,\varphi)$.
Then our geometrical analysis results in the following:
\begin{equation}
    Z(\KB\times S^1_C)=\tr{}_{V(T^2)}(P_{T^2}).\label{dimkb}
\end{equation}
Here $P_{T^2}$ is the reflection operator acting on the Hilbert space on $T^2$ and can be identified with $P$ appearing in \eqref{Pmat}. Using the explicit form of \eqref{Pmat}, $\tr_{V(T^2)}(P_{T^2})$ turns out to be the number of anyons which satisfy $a=\sP a$, which is compatible with \cite[Sec.~VII]{Barkeshli:2016mew}. Hereafter, we will denote $P_{T^2}$ by $P$. We will compute this explicitly in Sec.~4.1 and check the compatibility with other result.

\section{Basics of abelian anyons}

\subsection{Definitions of abelian anyons data}
Let us define the minimum data of the abelian anyons system. Here, we consider abelian bosonic systems, which are well-defined without specifying the spin structure on the manifold. Required data are below:
\begin{itemize}
  \item $\cA$ : the group of charges of anyons, which is finite and abelian.
  \item $\theta$ : the topological spin, which is a function $\cA \rightarrow U(1)$ such that it is non-degenerate, homogeneous and quadratic.
  \item $c$ : the chiral central charge $\in \bZ$ satisfying the Gauss sum constraint.
\end{itemize}

With data above, we define the braiding phase as follows:
\begin{equation}
    \begin{array}{rccc}
B\colon &\cA \times \cA                     &\longrightarrow& U(1)                     \\
        & \rotatebox{90}{$\in$}&               & \rotatebox{90}{$\in$} \\
        & (a,b)                    & \longmapsto   & \theta(a+b)\theta(a)^{-1}\theta(b)^{-1}
\end{array}.
\end{equation}
    Note that $\theta$ is called non-degenerate if $B$ is non-degenerate; it is called quadratic if $B$ is a bihomomorphism; and it is called homogeneous if
\begin{equation}
    \theta(na)=\theta(a)^{n^2}.
\end{equation}
The Gauss sum constraint is the following:
\begin{equation}
    \frac{1}{|\cA|^{1/2}}\sum_{a\in\cA}\theta(a) =
    e^{2\pi i c/8}.
\end{equation}

\subsection{Presentation of $SL(2,\bZ)$ and $GL(2,\bZ)$}
In this section, we recall the presentation of $SL(2,\bZ)$ and $GL(2,\bZ)$.
Let us take the convention \begin{equation}
\rS=\begin{pmatrix}
0 & -1\\
1 & 0
\end{pmatrix},\quad
\rT=\begin{pmatrix}
1 & 1 \\
0 & 1 
\end{pmatrix}, \quad
\rP=\begin{pmatrix}
1 & 0 \\
0 & -1
\end{pmatrix}.
\end{equation}

$SL(2,\bZ)$ has the following presentation. (See e.g. \cite[Appendix A]{KasselTuraev}): \begin{equation}
SL(2,\bZ)=\langle \rS,\rT,\rC \mid \rS^2=(\rS\rT)^3=\rC, \rC^2=1\rangle. \label{SL2Z}
\end{equation}

Elements of $GL(2,\bZ)$ is either $g$ or $\rP g$, where $g\in SL(2,\bZ)$.
And $g$ is always given by a string of ${\rS}^{\pm1}$ and ${\rT}^{\pm1}$.
Therefore, to fix the multiplication rules of $GL(2,\bZ)$,
we need to fix how the conjugation by $\rP$ acts on $\rS$ and $\rT$.
The necessary additional relations are \begin{equation}
{\rP}^2=1,\quad \rP\rS=\rC\rS\rP, \quad
\rP\rT={\rT}^{-1}\rP.
\end{equation} So we see that \begin{equation}
\begin{aligned}
GL(2,\bZ)=\langle \rS,\rT,\rC,\rP \mid& \rS^2=(\rS\rT)^3=\rC, \ \rC^2=1, \\
&\rP^2=1,\  \rP\rS=\rC\rS\rP,\ \rP\rT=\rT^{-1}\rP\rangle.
\end{aligned}\label{GL2Z}
\end{equation}

\subsection{Representation of $GL(2,\bZ)$ of abelian anyons}
In this section, we give explicit form of a $P$ matrix as a projective representation of $GL(2,\bZ)$. By the axioms, $V(T^2)$ is the representation space of $SL(2,\bZ)$ in the oriented cases, and the forms of the matrices are given as follows:
\begin{equation}
    S_{ab}=\frac{1}{|\cA|^{1/2}}B(a,-b), \quad T_{ab}=\delta_{ab}e^{-2\pi ic/24}\theta(a),\quad C_{ab}=\delta_{a,-b}.\label{STCmat}
\end{equation}

Now, let us move on to the non-orientable cases, where we need to consider $GL(2,\bZ)$.  We assume that the form of the $P_{ab}$ is given as follows:
\begin{equation}
    P_{ab} = \delta_{a,\sP b}\phi(b), \label{Pmatp}
\end{equation}
where $\phi(b)\in U(1)$ is a projective phase dependent on $b$.\footnote{We denote the elements of $SL(2,\bZ)$ or $GL(2,\bZ)$ as $\rS,\rT,\rC$ and $\rP$, the matrix representation of $SL(2,\bZ)$ or $GL(2,\bZ)$ as $S,T,C$ and $P$, and the reflection operation on anyons as $\sP$.} This assumption corresponds to the fact that the particle $a$ gets reflected to $\sP a$ when $P$ acts on $V(T^2)$.
By checking whether $P$ behave as a projective representation of $GL(2,\bZ)$, we derive several properties of $\theta$ and $B$, show that the factor $\phi$ is also independent of its entry, and find that we can construct the projective representation of $GL(2,\bZ)$.
From \eqref{GL2Z}, we have the following relations:
\begin{equation}
    \rT\rP\rT=\rP, \label{TPT}
\end{equation}
\begin{equation}
    \rS\rP\rS=\rP \label{SPS}.
\end{equation}
Therefore, we will see whether the following relations holds:
\begin{equation}
    (TPT)_{ab}=\alpha P_{ab}, \label{projTPT}
\end{equation}
\begin{equation}
    (SPS)_{ab}=\beta P_{ab}, \label{projSPS}.
\end{equation}
where $\alpha,\beta\in U(1)$.
First, we look into the relation above \eqref{projTPT}. Substituting the expressions \eqref{STCmat} and \eqref{Pmatp}, we have
\begin{equation}
\begin{aligned}
    (TPT)_{ab} &= \sum_{l,m\in\cA}\delta_{al}e^{-2\pi ic/24}\theta(a)\delta_{l,\sP m}\phi(m)\delta_{mb}e^{-2\pi ic/24}\theta(m)\\
               &= e^{-2\pi ic/12}\theta(a)\theta(\sP a)\delta_{a,\sP b}\phi(b),\\
    P_{ab} &= \delta_{a,\sP b}\phi(b).\label{comptpt}
\end{aligned}
\end{equation}
In order for \eqref{comptpt} to agree with \eqref{projTPT} as a projective representation, we have
\begin{equation}
    \theta(a)\theta(\sP a)=\text{const}. \in U(1) \quad\text{(independent of entries)}.
\end{equation}
With the fact that $\theta(0)=\theta(0)^{0^2}=1$, we can conclude that
\begin{equation}
    \theta(\sP a)=\theta(a)^{-1}
    \quad \text{for all }a\in\cA.\label{ptheta}
\end{equation}
This equation \eqref{ptheta} is the standard fact which is often assumed or derived from other assumptions.
Using this, we get how $P$ is related to the function $B$ as follows:
\begin{equation}
    \begin{aligned}
        B(\sP a,\sP b)&=\theta(\sP a+\sP b)\theta(\sP a)^{-1}\theta(\sP b)^{-1}\\
                &=\theta(a+b)^{-1}\theta(a)\theta(b)\\
                &=B(a,b)^{-1},\\
         B(a,\sP b)&=B(\sP a,{\sP}^2b)^{-1}\\
                &=B(\sP a,b)^{-1}.
    \end{aligned}
\end{equation}

Now, we are ready to look into another relation \eqref{SPS}. Substituting the expressions \eqref{STCmat} and \eqref{Pmatp}, we have
\begin{equation}
    \begin{aligned}
    (SPS)_{ab} &= \frac{1}{|\cA|}\sum_{l,m\in\cA}B(a,-l)\delta_{l,\sP m}\phi(m)B(m,-b)\\
               &= \frac{1}{|\cA|}\sum_{m\in\cA}B(a,-\sP m)B(m,-b)\phi(m)\\
               &= \frac{1}{|\cA|}\sum_{m\in\cA}B(\sP a,m)B(m,-b)\phi(m)\\
               &= \frac{1}{|\cA|}\sum_{m\in\cA}B(m,\sP a-b)\phi(m),\\
    P_{ab} &= \delta_{a,\sP b}\phi(b).
\end{aligned}
\end{equation}
Here, we set $b=\sP a$, then we have
\begin{equation}
    \frac{1}{|\cA|}\sum_{m\in\cA}\phi(m)/\phi(b)\in U(1).
\end{equation}
This holds true if and only if
\begin{equation}
    \phi(m)=\phi(0)\in U(1) \quad\text{(independent of entries)}, \label{trivphasep}
\end{equation}
where we use the triangle inequality.
Then we can absorb the constant factor $\phi(0)$ into the operator by redefining. Therefore we have
\begin{equation}
    P_{ab}=\delta_{a,\sP b} \label{Pmat}.
\end{equation}
One can check $S,T,C$ and $P$ above are compatible with other equations in \eqref{GL2Z} as a projective representation.\footnote{We can say that this is an ordinary representation rather than a projective one. Indeed, it is known that for a consistent quantum theory the central charge satisfies
\(c \equiv 0,12 \pmod{24}\)~\cite{Geiko:2022qjy}. As a consequence, the would-be projective phase in~\eqref{comptpt} is trivial, and the resulting action defines a genuine representation.
}
In this way, we can conclude that $V(T^2)$ is a projective representation space of $GL(2,\bZ)$. A brief summary of this section is below:
\begin{equation}
    \begin{aligned}
        S_{ab}&=\frac{1}{|\cA|^{1/2}}B(a,-b),\\
        T_{ab}&=\delta_{ab}e^{-2\pi ic/24}\theta(a),\\
        C_{ab}&=\delta_{a,-b},\\
        P_{ab}&= \delta_{a,\sP b},\\
        \theta(\sP a)&=\theta(a)^{-1},\\
         B(\sP a,\sP b)&=B(a,b)^{-1},\\
         B(a,\sP b)&=B(\sP a,b)^{-1}. \label{formulas}
    \end{aligned}
\end{equation}
Finally, let us relate the above results of the projective representation of $GL(2,\bZ)$ to the physics language in abelian bosonic systems.
\begin{itemize}
    \item $S$ : $S_{ab}$ can be viewed as the partition which is a link invariant given by the worldline of $a$ and $b$ embedded in $S^3$ since we have $S^3=(S^1\times D^2)\cup_{T^2}(D^2\times S^1)$ .\footnote{See \cite[Sec.~7]{Simon:2023hdq}.}
    \item $T$ : $T$ implements the Dehn twist acting on $T^2$. The action of the Dehn twist does not change the state on $T^2$ because it does not affect the geometry of $T^2$. But that may give us non-trivial phase factor which we identify as the topological spin.\footnote{See \cite[Sec.~17]{Simon:2023hdq}.}
    \item $C$ : $C$ implements the charge conjugation; $a\mapsto \overline{a}=-a$.
    \item $P$ : $P$ implements the reflection; $a\mapsto \sP a$.

\end{itemize}

\subsection{Other properties of anyons data}
In the later discussions, we will use several facts which were shown 
in \cite[Supplementary Materical, Sec.~B]{Wang_2017}
and also in \cite[Sec.~3.2]{Lee:2018eqa}.
Here we mostly follow the latter reference, but there is a slight difference in notation. We can show the same properties by redefining $\sT$ in \cite{Lee:2018eqa} to $-\sP$. The reason for the replacement is that from $\mathsf{CPT}$ invariance, we have $\sP a=\overline{\sT a}=-\sT a$ in abelian cases.

We start from the fact\footnote{In fact, one can derive the assumption i) from the assumption ii) using the non-degeneracy of $B$.} which we showed in Sec.~3.3.
\begin{enumerate}
  \item $\sP$ : $\cA \rightarrow \cA$ satisfying $\sP^2=1$,
    \item $\theta(\sP a)=\theta(a)^{-1}$.
\end{enumerate}

Then, as a result, we get the following properties.
\begin{enumerate}
   \item The group $\cC$ $\coloneqq$ $\Ker(1+\sP)/\Im(1-\sP)=(\bZ_2)^{2m}$ for some $m\in\bZ$.
   \item $\sP$ satisfies the following relations:
   \begin{equation}
\Ker (1+\sP) = [\Im (1-\sP)]^\perp,\qquad
\Ker (1-\sP) = [\Im (1+\sP)]^\perp.
\label{ddd}
\end{equation}
\end{enumerate}
 
Using the results above, we have
\begin{equation}
    |\Ker(1+\sP)|/|\Im(1-\sP)|=2^{2m} \label{C},
\end{equation}
\begin{equation}
     |\cA|=|\Im(1+\sP)||\Ker(1+\sP)| 
          =|\Im(1-\sP)|^2 2^{2m} \label{A}.
\end{equation}

\section{Derivation of $M_a$ and time-reversal anomaly}
\paragraph{Main result :}The main result of this section, or this paper, is the derivation of the form of $M_a$, which turns out to be an integer. The fact that $M_a$ is an integer makes it possible to cut and glue the surfaces compatibly and guarantees that the dimension of the Hilbert space on any non-orientable surface is an integer. In Sec.~4.1, we derive the form of $M_a$ explicitly. In Sec.~4.2 and Sec.~4.4, we derive some constraints on $\cM$, where we define $\cM$ as a subgroup of $\cA$ whose elements give us  nonzero value to $M_a$. In Sec.~4.3, we derive the consistency condition on $B$ and $\eta$ which can be found in \cite[Eq. (234)]{Barkeshli:2016mew}\cite[Sec.~2.4]{Lee:2018eqa}. 

\subsection{Expression of $M_a$}

Using the expression of \eqref{STCmat} in \eqref{maa}, $M_a$ is given as follows:
\begin{equation}
    \begin{aligned}
        M_a &= \frac{1}{|\cA|^{1/2}}\sum_{b\in\Ker(1+\sP)}B(a,-b)\eta(b)\\
        &=\frac{1}{|\cA|^{1/2}}\sum_{b\in\Ker(1+\sP)}B(-a,b)\eta(b)
        . \label{M_a}
    \end{aligned}
\end{equation}
Note that $\{B(a,b)\}_{b\in\Ker(1+\sP)}$ and $\{\eta(b)\}_{b\in\Ker(1+\sP)}$ both define the one-dimensional representations of $\Ker(1+\sP)$. Therefore, by the orthogonality of irreducible characters, we have
\begin{equation}
    M_a=\frac{|\Ker(1+\sP)|}{|\cA|^{1/2}}f(a),
\end{equation}
where we define the map $f:\cA\rightarrow\{0,1\}$ as follows:
\begin{equation}
f(a)=\begin{cases}
1 & (B(-a,b)\eta(b)=1\quad\text{for all }b\in\Ker(1+\sP))\\
0 & (\text{otherwise})
\end{cases}.
\end{equation} 
The discussion up to this point can also be found in \cite{Wang_2017}.

With (\ref{C}) and (\ref{A}), we have
\begin{equation}
    M_a = 2^m f(a).
\end{equation}

Now we define $\cM$ as follows:
\begin{equation}
    \cM\coloneqq\{a\in\cA \mid f(a)=1
    \}
\end{equation}
Then let us relate this expression of $M_a$ with our previous result.
Recall \eqref{dimkbb} and \eqref{dimkb}, we have
\begin{equation}
    \begin{aligned}
        \sum_{a\in\cA}M_aM_{-a}
        &=\tr{}_{V(T^2)}(P)\\
        &=\tr{}_{V(T^2)}(SPS^{-1})\\
        &=|\Ker(1+\sP)|\\&
    =\sum_{a\in\cM}2^{2m}f(a)f(-a)\\
        &=2^{2m}|\cM|,\\ \label{dimdim}
    \end{aligned}
\end{equation}
where we assume $f(-a)=1$ when $f(a)=1$. This assumption is necessary for cutting and gluing operations to be compatible, and we will prove it algebraically in \eqref{-a}. Also note that we change the bases in the second equality for the later convenience.
Then, we sum $M_a$ over $a$ with (\ref{M_a}), we have
\begin{equation}
    \begin{aligned}
            \sum_{a\in\cA}M_a&=2^m|\cM|\\
            &=\sum_{a\in\cA}\frac{1}{|\cA|^{1/2}}\sum_{b\in\Ker(1+\sP)}B(-a,b)\eta(b)\\
                         &=\frac{1}{|\cA|^{1/2}}\sum_{b\in\Ker(1+\sP)}\eta(b)\sum_{a\in\cA}B(-a,b)\\
                         &=\frac{1}{|\cA|^{1/2}}\sum_{b\in\Ker(1+\sP)}\eta(b)|\cA|\delta_{b,0}\\
                         &=|\cA|^{1/2}. \label{masum}
    \end{aligned}
\end{equation}
Therefore with \eqref{A}, \eqref{dimdim}, and \eqref{masum}, we have
\begin{equation}
\begin{aligned}
    2^{2m}&=\frac{|\Ker(1+\sP)|}{|\cM|}\\
    &=\frac{|\Ker(1+\sP)|}{2^{-m}|\cA|^{1/2}}\\
    &=\frac{|\Ker(1+\sP)|}{|\Im(1-\sP)|}\\
    &=|\cC|.
\end{aligned}
\end{equation}
This is compatible with the previous result \eqref{C}.
Please also note that 
\begin{equation}
    |\cM|=|\Im(1-\sP)| \label{dorder}
\end{equation}
which we will recall in Sec.~4.4.
 
\subsection{Constraints on $\cM$}
Let us derive some constraints on $\cM$ as folllows:
\begin{align}
    \sP a&\in\cM&\text{for all }a\in \cM,\label{apa}\\
    \theta(a) = \text{const}. &\in \{ \pm1\} &\text{for all } a\in \cM ,\label{consttheta}
\end{align}
In addition, define a subgroup of $\cA$ as
\begin{equation}
    \cM_-\coloneqq\{a\in\cA\mid\exists(b,c)\in\cM\times\cM,\quad a=b-c\}.
    \end{equation}
Then we will show
\begin{equation}
    \cM_-=\Im(1-\sP) .\label{im1-p}
\end{equation}
\paragraph{\eqref{apa} :}
We can paraphrase the fact that $a$ is the element of $\cM$ by
\begin{equation}
    B(-a,b)\eta(b)=1 \quad\text{for all }b\in \Ker(1+\sP). \label{ddef}
\end{equation}
Then, we have
\begin{equation}
    \begin{aligned}
        B(-\sP a,b)\eta(b)&=B(a,\sP b)\eta(b) \\
        &=B(a,-b)\eta(b)\\
                      &=B(-a,b)\eta(b)\\
                      &=1 \quad\quad\text{for all }b\in \Ker(1+\sP).
    \end{aligned}
\end{equation}

\paragraph{\eqref{consttheta} :}
The equation \eqref{consttheta} follows from the fact that the Dehn twist $T$ acting on $\MO_A\times S^1_B$ does not change the topology of the manifold. That means the resulting state sits in the same ray as the ordinary state in the Hilbert space, i.e. they are proportional to each other with some phase factor valued in $U(1)$:
\begin{equation}
    \begin{aligned}
    T\ket{\bC\bC}&=\sum_{a\in\cA }M_ae^{-2\pi ic/24}\theta(a)\ket{a}\\
&=\theta_{\bC\bC}\ket{\bC\bC}\quad\text{$\theta_{\bC\bC}\in U(1)$}.
    \end{aligned}
\end{equation}
This leads to the fact that every $\theta(a)$ where $a\in\cM$ has the same value. We denote this constant value as $\theta_\cM$ hereafter:
\begin{equation}
    \theta_\cM\coloneqq\theta(a) \quad\text{for all }a\in \cM.
\end{equation}
And recall the fact that $\theta(\sP a)=\theta(a)^{-1}$, we have
\begin{equation}
    \begin{aligned}
        \theta(a)\theta(\sP a)=\theta_\cM^2&=1,\\
        \text{i.e.} \quad \theta_\cM&=\pm1 .
    \end{aligned}
\end{equation}
This is considered one of the constraints to be satisfied for an anomaly-free theory\cite{Barkeshli:2017rzd}.

\paragraph{\eqref{im1-p} :}
First, we show $\cM_-\subset \Im(1-\sP)$.
Suppose we have $a\in\cM_-$, then there exists $(b,c)\in\cM\times\cM$ which satisfies $a=b-c$. Then we have
\begin{equation}
    \begin{aligned}
        B(-b,d)\eta(d)&=1,\\
        B(-c,d)\eta(d)&=1,\\
        \text{i.e.}\quad B(b-c,d)=B(a,d)&=1 \quad\text{for all } d\in \Ker(1+\sP).
    \end{aligned}
\end{equation}
Therefore the discussion in Sec.~3, \eqref{ddd} in particular, shows us that $a\in\Im(1-\sP)$.

Second, we show $\Im(1-\sP)\subset\cM_-$. Suppose we have $a\in\Im(1-\sP)$. Then take some $b\in\cM$ and consider $c=a+b$. Then we see that
\begin{equation}
    \begin{aligned}
        B(-c,d)\eta(d)&=B(-a-b,d)\eta(d)\\
        &=B(-a,d)B(-b,d)\eta(d)\\
        &=1 \quad\quad\quad\text{for all } d\in \Ker(1+\sP).
    \end{aligned}
\end{equation}
Therefore $c$ is also the element of $\cM$; i.e. we have $a=c-b\in\cM_-$.

\subsection{Consistency condition on $B$ and $\eta$}
In this section, using facts in Sec.~4.2, we will show
\begin{equation}
    B(a,-\sP a)\eta((1-\sP)a)=1 \quad\text{for all } a\in \cA.\label{consistency}
\end{equation}
This relation can be shown in terms of MTC and considered one of the nontrivial constraint on $\eta$ \cite[Eq. (234)]{Barkeshli:2016mew}\cite[Sec.~2.4]{Lee:2018eqa}. We verify that the same constraint is satisfied in our setup.

At first, we will derive
\begin{equation}
     B(a,-\sP a)\eta((1-\sP)a)=\frac{\theta(c+(1-\sP)a)\eta(c+(1-\sP)a)}{\theta(c)\eta(c)}\quad\text{for all }c\in\Ker(1+\sP), \text{ for all }a\in\cA. \label{wellB}
\end{equation}
This relation is shown as follows:
\begin{equation}
    \begin{aligned}
       \frac{\text{RHS}}{\text{LHS}}&= \frac{\theta(c+(1-\sP)a)\eta(c)\eta((1-\sP)a)\theta(a)\theta(\sP a)}{\theta(c)\eta(c)\theta((1-\sP)a)\eta((1-\sP)a)}\\
       &=\frac{\theta(c+(1-\sP)a)}{\theta(c)\theta((1-\sP)a)}\\
       &=B(c,(1-\sP)a)\\
       &=B((1+\sP)c,a)\\
       &=1.
    \end{aligned}
\end{equation}
Here, we used the fact that $\eta$ is a homomorphism and that $c\in\Ker(1+\sP)$.

Next, we move on to show that $\theta(a)\eta(a)$ is well-defined on $\cC$; the right-hand side of \eqref{wellB} is 1. 
Recall that every $a\in\cM$ has a constant value of the topological spin $\theta_\cM$. 
For any $a\in\cA$, there exists $(f,e)\in\cM\times\cM$ which satisfies $(1-\sP)a=f-e$ because of \eqref{im1-p}. Therefore we have
\begin{equation}
    \begin{aligned}
        1&=\frac{\theta(f)}{\theta(e)}\\
        &=\frac{\theta(e+(1-\sP)a)}{\theta(e)}\\
        &=B(e,(1-\sP)a)\theta((1-\sP)a)\\
        &=\eta((1-\sP)a)\theta((1-\sP)a)\\
    \end{aligned}\label{balbe}
\end{equation}
Here, at the fourth equality, we used
\begin{equation}
    \begin{aligned}
        B(-e,(1-\sP)a)\eta((1-\sP)a)&=1,\\
\text{i.e.}\quad B(e,(1-\sP)a)&=\eta((1-\sP)a)
    \end{aligned}
\end{equation}
since $e\in\cM$ and $(1-\sP)a\in\Ker(1+\sP)$. Using \eqref{balbe}, for all $c\in\Ker(1+\sP)$ and $a\in\cA$ we have
\begin{equation}
    \begin{aligned}
        \frac{\theta(c+(1-\sP)a)\eta(c+(1-\sP)a)}{\theta(c)\eta(c)}&=B(c,(1-\sP)
    a)\theta((1-\sP)a)\eta((1-\sP)a)\\
        &=B((1+\sP)c,a)\\
        &=1.
    \end{aligned}
\end{equation}
With \eqref{wellB}, we get the result as follows:
\begin{equation}
    B(a,-\sP a)\eta((1-\sP)a)=1 \quad\text{for all } a\in \cA. 
\end{equation}

Here, note that $B(a,-\sP a)\in\{\pm1\}$, therefore $\eta(a)\in\{\pm1\}$ for all $a\in\Im(1-\sP)$.
Also let us define a function $q:\cC\rightarrow U(1)$ as $q(a)=\theta(a)\eta(a)$. Then we have
\begin{equation}
\begin{aligned}
    q(a)^2&=q(2a)/B(a,a)\\
    &=q(0)B(a,a)\\
&=1\\
\text{i.e. }q(a)&\in\{\pm1\}\quad\text{for all }a\in\cC
\end{aligned}
\end{equation}
since $q$ is well-defined on $\cC=(\bZ_2)^{2m}$. Note that we have $\theta(a)\in\{\pm1\}$ for all $a\in\Ker(1+\sP)$ since $\theta(a)^2=\theta(a)\theta(-\sP a)=1$. Therefore, we have $\eta(a)\in\{\pm1\}$ for all $a\in\cC$.

Combining the two, we have
\begin{equation}
    \eta(a)\in\{\pm1\}\quad\text{for all } a\in\Ker(1+\sP),\label{etapm}
\end{equation}
since we can write any element of $\Ker(1+\sP)$ as $a=c+(1-\sP)b$ for some $c\in\cC$ and $b\in\cA$.
This is a standard fact, which is often assumed or derived from other assumptions.
With this fact, we can easily show
\begin{equation}
    \eta(a)=\eta(-a)=\eta(\sP a),
\end{equation}
where we use $\eta(0)=1$.


\subsection{More constraints on $\cM$}
Using \eqref{etapm}, we can put more constraints on $\cM$ as follows:
\begin{align}
     -a&\in\cM\quad\text{for all }a\in \cM,\label{-a}\\
    \cM&=\{e+x \mid x\in\Im(1-\sP)\},\label{formd}
\end{align}
where $e\in\cA$ satisfies $2e\in\Im(1-\sP)$ and $f(e)=1$.
\paragraph{\eqref{-a} :}
Similarly to \eqref{apa}, for all $a\in\cM$, we have
\begin{equation}
    \begin{aligned}
        B(a,b)\eta(b)&=B(-a,-b)\eta(-b)\\
        &=1\quad\text{for all } b\in\Ker(1+\sP),
    \end{aligned}
\end{equation}
where we use $\eta(b)=\eta(-b)$.

\paragraph{\eqref{formd} :}
Pick an $e\in \cM$.
From \eqref{-a}, $-e\in \cM$.
From \eqref{im1-p}, $2e=e-(-e)\in \Im(1-\sP)$.
Also from \eqref{im1-p}, any $e'\in \cM$ is given as $e+x$, where $x\in\Im (1-\sP)$.
Therefore $|\cM| \le |\Im(1-\sP)|$,
where the equality is satisfied only when the relation \eqref{formd} holds.
As we already showed that $|\Im(1-\sP)|=|\cM|$ in \eqref{dorder},
the equality follows.

\subsection{Relation to time-reversal anomaly}
In this section, we review the relation between $\theta_\cM$ and the time-reversal anomaly which is discussed in \cite[Eq.~(6)]{Wang_2017}\cite[Sec.~VII]{Barkeshli:2016mew}.
In general, the time-reversal anomaly of 2+1 dimensional abelian bosonic systems is characterized by the abelian bosonic SPT phases in 3+1 dimensional space-time with time-reversal symmetry. Such SPT phases have been argued to have $\bZ_2\times\bZ_2$ classification\cite{kapustin2014symmetryprotectedtopologicalphases},
 and can be distinguished by two signs. Those two signs correspond to two partition functions $Z(\RP^4)$ and $Z(\CP^2)$, which are computed in \cite[Sec.~VII]{Barkeshli:2016mew} in general 2+1 dimensional non-spin cases. They become
\begin{equation}
    \begin{aligned}
        Z(\RP^4)&=\frac{1}{|\cA|^{1/2}}\sum_{a\in\Ker(1+\sP)}\theta(a)\eta(a),\\
        Z(\CP^2)&=\frac{1}{|\cA|^{1/2}}\sum_{a\in\cA}\theta(a)=e^{2\pi ic/8}.
    \end{aligned}
\end{equation}
for abelian cases. Note that in the abelian bosonic case, we have
\begin{equation}
\begin{aligned}
    Z(\CP^2)&=\frac{1}{|\cA|^{1/2}}\sum_{a\in\cA}\theta(a)
    &=\frac{1}{|\cA|^{1/2}}\sum_{a\in\cA}\theta(\sP a)\\&=\frac{1}{|\cA|^{1/2}}\sum_{a\in\cA}\theta(a)^{-1}
    &=\frac{1}{|\cA|^{1/2}}\sum_{a\in\cA}\theta(a)^\ast\\
    &=Z(\CP^2)^\ast,
\end{aligned}
\end{equation}
where $\theta(a)^\ast$ is the complex conjugate of $\theta(a)$. Therefore, we see that $Z(\CP^2)\in\bR$; i.e. $c=0,4$ $\mathop{\text{mod}}8$.\footnote{This can be also concluded by a direct computation of the path integral \cite[Sec.~VII]{Barkeshli:2016mew}\cite[Sec.~III]{PhysRevResearch.3.023107}.}
We derive the relation between $\theta_\cM$ and these two partition functions as follows:
\begin{equation}
    \begin{aligned}
        Z(\RP^4)Z(\CP^2)&=\frac{1}{|\cA|^{1/2}}\sum_{a\in\Ker(1+\sP)}\theta(a)\eta(a)\frac{1}{|\cA|^{1/2}}\sum_{b\in\cA}\theta(b)\\
        &=\frac{1}{|\cA|}\sum_{a\in\cA}\theta(a)\eta(a)\sum_{b\in\cA}\theta(b)\\
        &=\frac{1}{|\cA|}\sum_{a,b\in\cA}\theta(a+b)B(a,-b)\eta(a)\\
        &=\frac{1}{|\cA|}\sum_{a,c\in\cA}\theta(c)B(a,a-c)\eta(a)\\
        &=\frac{1}{|\cA|}\sum_{c\in\cA}\theta(c)\sum_{a\in\Ker(1+\sP)}B(-c,a)\eta(a)\\
        &=\frac{1}{|\cA|^{1/2}}\sum_{c\in\cA}\theta(c)M_c\\
        &=\theta_{\cM}.
    \end{aligned}
\end{equation}
In the above computation, we used the following:
\begin{itemize}
\item $(a,b)\mapsto(a,c=a+b)$
    \item $\eta(a)=0$ for all $ a \notin\Ker(1+\sP)$,
    \item $B(a,a)=\theta(a)^2=\theta(a)\theta(\sP a)=1$ for all $ a\in\Ker(1+\sP)$,
    \item $\theta(a)=\theta_\cM$ for all $ a\in\cM$,
    \item $\sum_{a\in\cA}M_a=\sum_{a\in\cM}M_a=|\cA|^{1/2}$.
\end{itemize}

\section{Conclusion and Outlook}
We construct the way to interpret the system on torus with time-reversal symmetry as the projective representation space of $GL(2,\bZ)$ in abelian bosonic cases in Sec.~3.3. In the discussion, we derive several facts about the topological spin and the pairing $B$ associated with the reflection. Also, we explicitly write down the expression of the dimension of the Hilbert space on $\RP^2$ with a line operator inserted in abelian bosonic cases in Sec.~4.1. The dimension results in $2^{m}$ or $0$. This enables us to compute the dimension of the Hilbert spaces compatibly on any (2+1)D non-orientable surface. This result is also important from the MTC point of view since it is considered one of the condition for a theory to be anomaly-free. Also in Sec.~4.3, we check that consistency condition are met with respect to $B$ and $\eta$ which is also needed to be anomaly-free. In Sec.~4.4 we give an explicit form of $\cM$ which gives a positive value to the dimension of $\RP^2$.
 In Sec.~4.5, we relate our result on the topological spin to the anomaly formula.

Our future work is to extend our discussion to non-abelian cases or spin cases. We hope to obtain a clearer understanding of the correspondence between Atiyah TQFT and MTC even in non-orientable cases.

\paragraph{Acknowledgements:} The author would like to thank Y. Tachikawa for helpful discussions and the anonymous referee for constructive comments.The author is supported in part by Forefront Physics and Mathematics Program to Drive Transformation (FoPM), a World-leading Innovative Graduate Study (WINGS) Program, at the University of Tokyo.

\bibliographystyle{ytamsalpha}
\baselineskip=.95\baselineskip
 \def\arxivfont{\rm}
\bibliography{ref}


\end{document}